\documentclass[10pt]{article}
\usepackage[margin=1in]{geometry}
\usepackage{palatino}
\usepackage{amsthm,amsmath,amssymb,amsfonts}
\usepackage[final]{graphicx}
\usepackage{subcaption}
\usepackage{hyperref}
\usepackage{booktabs}
\usepackage{algorithm, algpseudocode}
\usepackage{xcolor}
\usepackage{ifdraft}
\usepackage[utf8]{inputenc}
\usepackage{xspace}
\usepackage[final]{pdfpages}
\usepackage{graphicx}
\usepackage{verbatim}
\usepackage{xfrac}
\usepackage{array}

\newcommand{\dhcre}{\textsf{DHC-RE}\xspace}
\newcommand{\safetabp}{SafeTab\xspace}

\theoremstyle{definition}
\newtheorem{definition}{Definition}

\theoremstyle{plain}
\newtheorem{theorem}{Theorem}
\newtheorem{lemma}{Lemma}
\newtheorem{corollary}{Corollary}
\newtheorem{proposition}{Proposition}

\newcommand{\am}[1]{\ifdraft{\textcolor{magenta}{AM: #1}}\xspace}

\newcommand{\eat}[1]{}

\newcommand{\epsfrac}{\gamma}

\newcommand{\dd}{\,\mathrm{d}}

\graphicspath{{screenshots/}}

\title{Differentially Private Algorithms for 2020 Census Detailed DHC Race \& Ethnicity}
\author{Sam Haney, 
William Sexton, 
Ashwin Machanavajjhala, 
Michael Hay,
Gerome Miklau\\
Tumult Labs\thanks{Ashwin Machanavajjhala, Michael Hay and Gerome Miklau are faculty members at Duke University, Colgate University and University of Massachussets, Amherst, respectively. Work done when authors were at Tumult Labs.}, 
\url{science@tmlt.io}
}
\date{}
\begin{document}

\maketitle

\begin{abstract}
This article describes a proposed differentially private (DP) algorithms that the US Census Bureau is considering to release the Detailed Demographic and Housing Characteristics (DHC) Race \& Ethnicity tabulations as part of the 2020 Census. The tabulations contain statistics (counts) of demographic and housing characteristics of the entire population of the US crossed with detailed races and tribes at varying levels of geography. We describe two differentially private algorithmic strategies, one based on adding noise drawn from a two-sided Geometric distribution that satisfies "pure"-DP, and another based on adding noise from a Discrete Gaussian distribution that satisfied a well studied variant of differential privacy, called Zero Concentrated Differential Privacy (zCDP). We analytically estimate the privacy loss parameters ensured by the two algorithms for comparable levels of error introduced in the statistics. 
\end{abstract}

\section{Problem \& Desiderata}
\label{sec:problem}
The SafeTab algorithm produces differentially private tables of statistics (counts) of demographic and housing characteristics of all persons in the US crossed with detailed races and tribes at varying levels of geography (national, state, county, AIANNH). In this section, we define relevant concepts, outline the statistics to be released, and then formulate the differentially private algorithm design problem. 

\paragraph{Definitions:}
Every person resides in exactly one census block that determines their geographic location. This Census block is contained in several \textit{geographic entities} -- e.g. LA county, the state of CA and the US. We assume there are hierarchical relationships between geographical entities. 

Every person is also associated with one or more  \textit{race codes} and an \textit{ethnicity code}. The maximum number of race codes a person can be associated with is called the \textit{race multiplicity}. A \textit{race (ethnicity) group} is a set of race (ethnicity) codes. An individual person is in a race group \textit{Alone} if all race codes associated with that individual are contained in the race group. A record is in a race group \textit{Alone or in Combination} if some race code associated with that record is contained in the race group.

A \textit{characteristic iteration} is the combination of a race (or ethnicity) group, along with the specification of either Alone or Alone or in Combination. A Characteristic Iteration has a corresponding Characteristic Iteration Code. (E.g. Latin American Indian (6800-6999) Alone or in Combination is a characteristic iteration, where 6800-6999 is the range of codes codes in the corresponding race group). Like geographical entities, characteristic iterations also have hierarchical relationships. 
One person may correspond to multiple characteristic iterations.

\paragraph{Detailed DHC (Race \& Ethnicity) Data Product:}
The Detailed DHC (Race \& Ethnicity) data product (\dhcre) aims to tabulate statistics by population groups. A \textit{population group} is a pair $(g, c)$, where $g$ is a geographic entity (e.g., the state of NC, or LA County) and $c$ is a race or ethnicity characteristic iteration (e.g., Latin American Indian (6800-6999) Alone or in Combination). 

Two different statistics are under consideration to be released for each population group. Given as input a dataframe of all US persons and their attributes, output:
\begin{itemize}
\item[(T1)]  total population associated with each population group. 
\item[(T2)]  Sex $\times$ Age marginal for a subset of population groups (that are not marked as \textit{TotalOnly}). 
\end{itemize}

\paragraph{Private Release Problem:}
The release of statistical data products by the US Census Bureau about persons and households is regulated under Title 13 and any release of statistics about persons in the US must be afforded strong privacy protections \cite{title13}. Moreover, it has been demonstrated that legacy statistical disclosure limitation (SDL) techniques are vulnerable to attacks that can reconstruct the sensitive person records from aggregate statistics \cite{ap-census-attack}. Hence, the US Census Bureau has decided that all statistics released as part of the 2020 Census (of which \dhcre is a part) will be released using algorithms that satisfy modern privacy definitions like differential privacy \cite{dsep-dp}.

In this paper, we describe SafeTab, a candidate differentially private algorithm for releasing the T1 and T2 statistics that make up a portion of the \dhcre data product. 

Based on workshops and in-depth discussions\footnote{The methodology and tools used to elicit preferences of users on the statistics to be released, privacy parameters, accuracy constraints, etc. is out of scope for this paper and will be the focus of a separate paper.} with users of the \dhcre data product and US Census Bureau data stewards, the following desiderata were identified for the differential privacy algorithm: 
\begin{itemize}
    \item \textit{Privacy:} The algorithm must ensure end-to-end differential privacy with respect to (the addition/removal of) every person in the US. 
    \item \textit{Population Groups:} The algorithm must release statistics for a predefined set of race and ethnicity characteristic iterations and following geographies: national level (US), 50 states + DC, counties within the 50 states + DC and all areas designated as American Indian Alaska Native and Native Hawaian (AIANNH) areas. \am{It would be good to distinguish between detailed and regional CIs here.} 
    \item \textit{Adaptivity:} The algorithm may adaptively choose the granularity at which Sex $\times$ Age statistics are released. For instance, for population groups with a few people the Sex $\times$ Age histogram may only have 4 buckets of age, while for population groups with many people a more detailed histogram may be released. 
    \item \textit{Accuracy:} Subject matter experts provided candidate accuracy levels for population groups in terms of the margin of error (MOE) in output counts. Different population groups had different MOEs specified (described later in the paper in Table~\ref{tab:moe-targets}). 
    \item \textit{Integrality:} The output statistics must be integral. 
    \item \textit{No Consistency:} The SafeTab differential privacy algorithm does not enforce consistency of any form. That is, different counts output by the systems need not be consistent with each other (e.g., the number of people of a certain characteristic iteration in the US need not equal the sum of the population counts for the same characteristic iterations across all states). The counts may also be negative. 
\end{itemize}
In the rest of the paper, we describe the SafeTab differential privacy algorithms and analyze bounds on the privacy loss achievable while satisfying the constraints mentioned above.

\section{SafeTab Algorithm}
\label{sec:algorithm-description}
\safetabp is a candidate privacy algorithm for releasing detailed race and ethnicity statistics from the 2020 Census. The algorithm accommodates the release of tabulations for total counts by detailed race and ethnicity and tabulations for sex by age counts by detailed race and ethnicity. The algorithm acts on a private dataframe derived from the 2020 Census. There is a row for each person in the US with attributes for which census block the individual resides in, race and ethnicity codes, sex, and age.

In this section we present an analysis of a simplified version of the algorithm. In particular, SafeTab produces tabulations at the level of a \emph{population group}. In reality, a population group is a geographic entity (e.g. a specific county) and a characteristic iteration code (see Section~\ref{sec:problem} for more details). Records are associated with population groups via algorithms that map their block id  to geographic entities, and their race and ethnicity codes to iteration codes. Additionally, population groups are split into levels (both geography levels and iteration levels) with distinct privacy loss budgets. For the purposes of this section, we assume the following model for population groups:

\begin{itemize}
    \item SafeTab should produce tabulations on sets of population groups $\mathcal{P}_1,\ldots,\mathcal{P}_{\omega}$, which we call \emph{population group levels}. That is, it should produce a tabulation for each population group $P \in \mathcal{P}_i$ for $1 \le i \le {\omega}$.
    \item SafeTab is provided  privacy loss budgets for each population group level $\rho_1, \ldots, \rho_{\omega}$ with $\rho_i$ corresponding to the budget for population group level $\mathcal{P}_i$.
    \item For each $\mathcal{P}_i$, we assume we have a function $g_i: \mathcal{I} \rightarrow 2^{\mathcal{P}_i}$, where $\mathcal{I}$ is the domain of records in the private dataframe. That is, $g_i$ maps a record to the subset of population groups at level $i$ to which it belongs.
    \item We assume the stability of $g_i$, denoted by $\Delta(g_i)$ is known. The stability is defined as $\Delta(g_i) = \max_{r \in \mathcal{I}} |g_i(r)|$.
\end{itemize}

The main algorithm is presented in Algorithm~\ref{alg:safetab-main-algorithm}.
This algorithm proceeds by looping over the population group levels.
For each population group level, we apply $g_i$ to the dataframe to map each record to the set of population groups it is associated with.
Then for each population group in the level, we call the tabulation function \textsc{TabulatePopulationGroup}, passing in a dataframe containing just the records in that population group.

The pseudocode for the procedure \textsc{TabulatePopulationGroup} is given in Algorithm~\ref{alg:safetab-tabulate-pop-group}. This code tabulates a single population group.
Population groups are characterized based on the tabulation we would like to compute. In particular, we assume we are given a set $TotalOnly$ of population groups for which only the size of the population group should be tabulated. We check whether the given group is a member of this set.
If it is, we call the \textsc{NoisyCount} function on the population group, which tabulates a noisy count of the size of the group.
Otherwise, we use a two stage algorithm.
We first compute a noisy count of the group using \textsc{NoisyCount}, but using only a fraction (denoted $\epsfrac$) of the available privacy loss budget. Next, we compare this noisy count against a set of given thresholds, denoted $\Theta_1, \Theta_2,$ and $\Theta_3$. Depending on which thresholds the noisy count exceeds, we compute sex by age noisy counts with a varying degree of age bin sizes. Age bins are coarser for smaller noisy counts. These sex by age counts are also computed by \textsc{NoisyCount} using the remaining privacy loss budget.

The pseudocode for the procedure \textsc{NoisyCount} is given in Algorithm~\ref{alg:noisy-count}. This procedure computes the number of rows in the dataframe and adds noise from either the discrete Gaussian distribution (see Algorithm~\ref{alg:base-discrete-gaussian}) or the two-sided geometric distribution (see Algorithm~\ref{alg:base-geometric}), depending on the value of $\Gamma$. The parameter $\rho$ can be interpreted as the pure DP loss (when $\Gamma$ is \emph{Geometric}) or as the zCDP loss (when $\Gamma$ is \emph{Discrete Gaussian}).

The notation used in this section and the algorithm pseudocode is summarized in Table~\ref{tab:algorithm-notation}. 

\section{Privacy \& Error Analysis}
\paragraph{Privacy Analysis}
\label{sec:tpdp-privacy-analysis}

We analyze the privacy loss of the SafeTab algorithm employing two different base privacy mechanisms: the geometric mechanism and the discrete Gaussian mechanism. The geometric mechanism is known to satisfy pure $\epsilon$-differential privacy whereas the discrete Gaussian mechanism satisfies $\rho$-zCDP. We utilize composition rules in conjunction with the privacy properties of the base mechanisms to derive privacy guarantees for the SafeTab algorithm under the two different mechanisms. To do so, we establish a novel generalized parallel composition result (Section \ref{sec:generalized-parallel-composition}). When running a set of $\epsilon$-differentially private mechanisms on a family of subsets $F = \{S_i\}$ of the input domain, the privacy-loss scales linearly with the maximum  number of sets containing any fixed element of the input domain. A similar result holds for zCDP. Applying the generalized composition results along with standard sequential composition rules, we prove SafeTab[Geometric] satisfies pure $\epsilon$-differential privacy and SafeTab[Discrete Gaussian] satisfies $\rho$-zCDP. We also analyze SafeTab[Geometric] under another variant of differential privacy known as R\'enyi differential privacy (RDP). See Appendix \ref{sec:safetab-geometric-privacy} and \ref{sec:safetab-discrete-gaussian-privacy} for privacy analysis details.   

\paragraph{Error Analysis}

To compare the privacy loss of SafeTab[Geometric] and SafeTab[Discrete Gaussian], we convert the privacy guarantees outlined in Section \ref{sec:tpdp-privacy-analysis} to approximate differential privacy where $\delta$ is fixed at $10^{-10}$. We set SafeTab parameter values based on input from subject matter experts at Census (see Appendix \ref{sec:compare} for details). Among the fixed parameters, are candidate target 95\% margin of errors. For SafeTab[Geometric], we calibrate the privacy loss parameter under pure differential privacy to achieve the target MOEs. Likewise, for SafeTab[Discrete Gaussian], we calibrate the privacy loss parameter under zCDP to achieve the target MOEs. Table \ref{tab:tpdp-moe-targets} illustrated the privacy loss parameters needed to guarantee desired MOEs. 

{\small
\begin{table}[t]
    \centering
    \begin{tabular}{c c c c c c}
    \toprule
    Population Group Level & MOE Target &  
    \multicolumn{2}{c}{Geometric ($\epsilon$)} & \multicolumn{2}{c}{Discrete Gaussian ($\rho$)}
    \\
    \cmidrule(lr{.75em}){3-4}
    \cmidrule(lr{.75em}){5-6}
    & & Step 2 & Total & Step 2 & Total \\
    \midrule
    (Nation, Detailed) & 6 & 3.84 & 4.27 & 0.481 & 0.534\\
    (State, Detailed) & 6 & 3.84 & 4.27 & 0.481 & 0.534\\
    (County, Detailed) & 11 & 2.24 & 2.49 & 0.143 & 0.159 \\
    (AIANNH, Detailed) & 11 & 2.24 & 2.49 & 0.143 &0.159\\
    (Nation, Regional) & 50 & 0.531 & 0.59 & 0.007 & 0.008\\
    (State, Regional) & 50 & 0.531 & 0.59 & 0.007 &0.008\\
    (County, Regional) & 50 & 0.531 & 0.59 & 0.007 &0.008\\
    \bottomrule
    \end{tabular}
   \caption{MOE targets for the statistics released (in Step 2 of the adaptive algorithm) at different population group levels along with the corresponding privacy loss ($\epsilon$-DP for Geometric and $\rho$-zCDP for Discrete Gaussian). The privacy loss is reported for the Step 2 (to match the MOE) as well as the total loss for that level. Step 2 loss is 90\% of Total loss at each population group level. The privacy losses reported here have already been aggregated over all the population groups at the given level, so the \emph{Total} column represents the privacy loss input parameters of the SafeTab algorithm.}
   \label{tab:tpdp-moe-targets}
\end{table}
}

Then, we covert the established total privacy loss values from table \ref{tab:tpdp-moe-targets} to approximate DP using conversion techniques outlined in Appendix \ref{sec:privacy-prelim}. For SafeTab[Geometric], we consider the total loss under pure DP ($\delta = 0$) as well as the approximate DP loss derived by first analyzing the geometric mechanism under RDP and then converting to approximate DP. For SafeTab[Discrete Gaussian], we convert the total loss under zCDP from Table \ref{tab:tpdp-moe-targets} to its equivalent loss under approximate differential privacy using two methods. One method uses an analytic derivation which is not tight, and the second method uses a numerical approximation to minimize equation \ref{equ:zcdp-to-approx-dp-tight} with respect to $\alpha \in (1, \infty)$.
{\small
\begin{table}
  \centering
  \begin{tabular}{r l l l l}
    \toprule
    \textbf{Base Mechanism} & \multicolumn{2}{c}{Geometric} & \multicolumn{2}{c}{Discrete Gaussian} \\
    \cmidrule(lr){2-3} \cmidrule(lr){4-5}
    \textbf{Analysis} & Pure DP & RDP & zCDP (analytical) & zCDP (experimental) \\
    \midrule
    \textbf{$(\epsilon, 10^{-10})$ privacy loss} & 15.3 ($\delta = 0$) & 13.2 & 12.8 & 12.2 \\
    \bottomrule
  \end{tabular}
  \caption{The approximate differential privacy loss of the SafeTab[Geometric] and SafeTab[Discrete Gaussian] algorithm for various analyses. For all but the Pure DP, $\delta=10^{-10}$. For the Pure DP analysis, $\delta=0$.}
  \label{tab:tpdp-privacy-loss-results}
\end{table}
}
\paragraph{Privacy losses of Geometric vs Discrete Gaussian:}
Results appear in Table~\ref{tab:tpdp-privacy-loss-results}. We observe the following key findings: First, as expected SafeTab[Geometric] permits a smaller privacy loss ($\epsilon$) under approximate DP with ($\delta = 10^{-10}$). We are able to achieve this by analyzing SafeTab[Geometric] under R\'enyi DP. Second, we observe that the privacy loss of SafeTab[Discrete Gaussian] is smaller than that of SafeTab[Geometric] (when $\delta=10^{-10}$). The improvement in privacy loss is small (3\% for the analytical bound and 7\% for the experimental bound).

{\small 
\bibliographystyle{plain}
\bibliography{refs}
}

\clearpage
\appendix

\section{Privacy Preliminaries}
\label{sec:privacy-prelim}

In this section, we give necessary background on differential privacy and related definitions of privacy.
In particular, we will analyze SafeTab using pure and approximate differential privacy, and zCDP, as well as with two different basic noise mechanisms, the geometric mechanism and the discrete Gaussian mechanism.

\subsection{Privacy definitions}
\label{sec:privacy-defintions}

\begin{definition}[Neighboring Databases]
Let $x,x'$ be databases represented as multisets of tuples. We say that $x$ and $x'$ are \emph{neighbors} if their symmetric difference is 1.
\end{definition}

We first define diffential privacy, the most common formal privacy definition.

\begin{definition}
An algorithm $M: \mathcal{X} \rightarrow \mathcal{Y}$ satisfies ($\epsilon, \delta$)-differential privacy if for all neighboring databases $x, x'$ and all output $y \in \mathcal{Y}$,
\begin{equation}
    P[M(x)=y] \leq e^\epsilon P[M(x')=y] + \delta
\end{equation}
\end{definition}

When a mechanism satisfies differential privacy with $\delta=0$, we say that the mechanism satisfies \emph{pure differential privacy}, and when $\delta>0$ we say the mechanism satisfies \emph{approximate differential privacy}.

We next define zCDP, which bounds the \emph{R\'enyi divergence} between between the distributions of a mechanism run on neighboring databases.

\begin{definition}
The \emph{R\'enyi divergence of order $\alpha$} between distribution $P$ and distribution $Q$, denoted $D_{\alpha}(P \| Q)$ is defined as
\begin{equation}
    D_{\alpha}(P \| Q) = \frac{1}{\alpha-1}\log\left(\underset{x \sim P}{\mathbb{E}} \left[ \left( \frac{P(x)}{Q(x)} \right)^{\alpha-1}\right]\right)
\end{equation}
\end{definition}

\begin{definition}(zCDP \cite{BunS16})
An algorithm $M: \mathcal{X} \rightarrow \mathcal{Y}$ satisfies $\rho$-zCDP if for all neighboring $x, x' \in \mathcal{X}$ and for all $\alpha \in (1, \infty)$,
\begin{equation}
    D_{\alpha}(M(x) \| M(x')) \le \rho \alpha.
\end{equation}
\end{definition}

Finally, we define R\'enyi differential privacy (RDP). RDP is similar to zCDP except that it (1) bounds the R\'enyi divergence of each order separately, and (2) allows for an arbitrary bound on the divergence, rather than requiring a bound that is linear in $\alpha$.

\begin{definition}[RDP \cite{Mironov17}]
\label{def:renyi-dp}
An algorithm $M: \mathcal{X} \rightarrow \mathcal{Y}$ satisfies $(\alpha, \epsilon)$-R\'enyi differential privacy ($(\alpha, \epsilon)$-RDP) if for all neighboring $x, x' \in \mathcal{X}$,
\begin{equation}
    D_{\alpha}(M(x) \| M(x')) \le \epsilon.
\end{equation}
\end{definition}

\subsection{Composition}
\label{sec:composition}

One of the most useful and important properties of privacy definitions is their behaviour under composition.
In this section, we state composition results for pure differential privacy, approximate differential privacy, zCDP, and RDP.
There are two types of composition we are interested in -- sequential composition and parallel composition. We first state the sequential composition results.

\begin{lemma}(Adaptive sequential composition of pure differential privacy)
\label{lem:sequential-composition-dp}
Let $M_1: \mathcal{X} \rightarrow \mathcal{Y}$ and $M_2: \mathcal{X} \times \mathcal{Y} \rightarrow \mathcal{Z}$ be mechanisms satisfying $\epsilon_1$-differential privacy and $\epsilon_2$-differential privacy respectively. Let $M_3(x) = M_2(x, M_1(x))$. Then $M_3$ satisfies $(\epsilon_1 + \epsilon_2)$-differential privacy.
\end{lemma}

\begin{lemma}(Adaptive sequential composition of zCDP \cite{BunS16})
\label{lem:sequential-composition-zcdp}
Let $M_1: \mathcal{X} \rightarrow \mathcal{Y}$ and $M_2: \mathcal{X} \times \mathcal{Y} \rightarrow \mathcal{Z}$ be mechanisms satisfying $\rho_1$-zCDP and $\rho_2$-zCDP respectively. Let $M_3(x) = M_2(x, M_1(x))$. Then $M_3$ satisfies $(\rho_1 + \rho_2)$-zCDP.
\end{lemma}

\begin{lemma}(Adaptive sequential composition of RDP \cite{Mironov17})
\label{lem:sequential-composition-rdp}
Let $M_1: \mathcal{X} \rightarrow \mathcal{Y}$ and $M_2: \mathcal{X} \times \mathcal{Y} \rightarrow \mathcal{Z}$ be mechanisms satisfying $(\alpha, \epsilon_1)$-RDP and $(\alpha, \epsilon_2)$-RDP respectively. Let $M_3(x) = M_2(x, M_1(x))$. Then $M_3$ satisfies $(\alpha, \epsilon_1 + \epsilon_2)$-RDP.
\end{lemma}

In Section~\ref{sec:generalized-parallel-composition} we state and prove generalized parallel composition lemmas for our privacy definitions.

\subsection{Converting zCDP and RDP to approximate differential privacy}

\begin{lemma}(\cite{CanonneK2020})
\label{lem:renyi-divergence-to-approx-dp-tight}
Let $M: \mathcal{X} \rightarrow \mathcal{Y}$ be a randomized algorithm and let $\alpha \in (1, \infty)$ and $\epsilon > 0$. Suppose $D_\alpha(M(x) \| M(x')) \le \tau$ for all neighboring $x,x'$. Then $M$ satisfies $(\epsilon, \delta)$-differential privacy where
\begin{equation}
    \delta = \frac{\exp((\alpha-1)(\tau-\epsilon))}{\alpha-1}\left(1 - \frac{1}{\alpha}\right)^{\alpha},
\end{equation}
or equivalently,
\begin{equation}
\label{equ:zcdp-to-approx-dp-tight}
\epsilon = \tau + \frac{\log(1/\delta) + (\alpha-1)\log(1 - 1/\alpha) - \log(\alpha)}{\alpha-1}.
\end{equation}
\end{lemma}

Lemma~\ref{lem:renyi-divergence-to-approx-dp-tight} can be used to convert both RDP and zCDP guarantees to approximate differential privacy guarantees.
For RDP, the conversion follows immediately from the lemma ($\tau$ in the lemma above the is the RDP $\epsilon$ parameter, which is different from the approximate differential privacy $\epsilon$ parameter).
Converting zCDP to approximate differential privacy using Lemma~\ref{lem:renyi-divergence-to-approx-dp-tight} requires minimizing over all possible values of $\alpha$ to find the tightest guarantee.
That is, if $M$ satisfies $\rho$-zCDP then it satisfies $(\epsilon, \delta)$-differential privacy where
\begin{equation}
    \epsilon = \inf_{\alpha \in (1,\infty)}\left[ \rho\alpha + \frac{\log(1/\delta) + (\alpha-1)\log(1 - 1/\alpha) - \log(\alpha)}{\alpha-1} \right].
\end{equation}
Computing this expression for $\epsilon$ can be done experimentally, but cannot easily be done analytically.
Because of this, we also give a looser conversion from zCDP to approximate differential privacy \cite{BunS16}.

\begin{lemma}(\cite{BunS16})
\label{lem:zcdp-to-approx-dp-loose}
Let $M: \mathcal{X} \rightarrow \mathcal{Y}$ be a randomized algorithm satisfying $\rho$-zCDP. Then for all $\delta>0$, $M$ satisfies $(\epsilon, \delta)$-differential privacy where
\begin{equation}
    \epsilon = \rho + \sqrt{4\rho\log(1/\delta)}.
\end{equation}
\end{lemma}

\subsection{Base Mechanisms}
\label{sec:base-mechansisms}

\begin{definition}
\label{def:discrete-gaussian-distribution}
The discrete gaussian distribution $\mathcal{N}_{\mathbb{Z}}(\sigma^2)$ centered at 0 is
\begin{equation}
\forall x \in \mathbb{Z}, \quad \Pr[X=x] = \frac{e^{-x^2/2\sigma^2}}{\sum_{y \in \mathbb{Z}}e^{-y^2/2 \sigma^2}}.
\end{equation}
\end{definition}

\begin{definition}
\label{def:geometric-distribution}
The two sided geometric (or discrete Laplace) distribution $\mathcal{L}_{\mathbb{Z}}(b)$ centered at 0 is

\begin{equation}
\forall x \in \mathbb{Z}, \quad \Pr[X=x]=\frac{e^{1 / b}-1}{e^{1 / b}+1} \cdot e^{-|x| / b}.
\end{equation}
\end{definition}

\begin{lemma}
\label{lem:geometric-satisfies-pure-dp}
Let $q: \mathcal{X} \rightarrow \mathbb{R}$. Then \textsc{BaseGeometric}$(q(x), \epsilon)$ from Algorithm~\ref{alg:base-geometric} satisfies $\epsilon$-differential privacy with respect to $x$.
\end{lemma}

\begin{lemma}{\cite{CanonneK2020}}
\label{lem:discrete-gaussian-satisfies-zcdp}
Let $q: \mathcal{X} \rightarrow \mathbb{R}$. Then \textsc{BaseDiscreteGaussian}$(q(x), \rho)$ from Algorithm~\ref{alg:base-discrete-gaussian} satisfies $\rho$-zCDP with respect to $x$.
\end{lemma}

Note that
\begin{itemize}
    \item \textsc{BaseDiscreteGaussian} does not satisfy pure differential privacy for any value of $\epsilon$.
    \item Any $\epsilon$-differentially private algorithm also satisfies $(\frac{1}{2} \epsilon^2)$-zCDP. There, \textsc{BaseGeometric}$(q(x), \epsilon)$ satisfies $(\frac{1}{2} \epsilon^2)$-zCDP.
    \item Analyzing \textsc{BaseGeometric}-based algorithms using $zCDP$ leads to looser bounds than using an RDP analysis. Therefore, in this paper we only consider an RDP analysis. The RDP guarantee for \textsc{BaseGeometric} is given in Corollary~\ref{cor:geometric-satisfies-renyi-dp}.
\end{itemize}

\section{Novel Privacy Results}
\label{sec:new-privacy}


In this section, we give some useful privacy results necessary for deriving the privacy losses of SafeTab.
To the best of our knowledge, these results are novel.

\subsection{R\'enyi parameters for base geometric}

We show that the base geometric mechanism satisfies R\'enyi differential privacy and give the R\`enyi privacy parameters for which this is true.
This will be useful for finding an approximate differential privacy guarantee for the composition of the geometric mechanism.
Lemma~\ref{lem:divergence-of-geometric-distribution} is an analogue of Proposition 6 in \cite{Mironov17}, and the proof is similar.

\begin{lemma}
\label{lem:divergence-of-geometric-distribution}
For any $\alpha > 1$ and $b > 0$,
\begin{equation}
D_{\alpha}(\mathcal{L}_{\mathbb{Z}}(b) \| (\mathcal{L}_{\mathbb{Z}}(b)+1)) = \frac{1}{\alpha-1} \log\left[\frac{e^{1 / b}-1}{e^{1 / b}+1}\left(\frac{2\alpha b}{2\alpha-1}e^{(\alpha-1)/b} + \frac{2\alpha b - 2b}{2\alpha-1}e^{-\alpha/b}\right)\right]
\end{equation}
\end{lemma}
\begin{proof}
Let $P$ and $Q$ be distributions with densities $p(x)$ and $q(x)$ respectively.
Then,
\begin{equation}
    D_{\alpha}(P \| Q) = \frac{1}{\alpha-1} \log{\int_{-\infty}^{\infty} p(x)^{\alpha}q(x)^{1-\alpha} \dd x}.
\end{equation}
For us,
\begin{equation}
    p(x) = \frac{e^{1 / b}-1}{e^{1 / b}+1} \cdot e^{-|x| / b}, \quad q(x) = \frac{e^{1 / b}-1}{e^{1 / b}+1} \cdot e^{-|x-1| / b}.
\end{equation}
We evaluate the integral separately on $(-\infty, 0)$, $(0,1)$, and $(1, \infty)$.
That is,
\begin{align}
\int_{-\infty}^{\infty} p(x)^{\alpha}q(x)^{1-\alpha} \dd x &=
\begin{aligned}[t]
\frac{e^{1 / b}-1}{e^{1 / b}+1} \cdot \Bigg[&\int_{-\infty}^{0}  \exp\left(\frac{\alpha x}{b} + \frac{(1-\alpha)(x-1)}{b}\right) \dd x \\
&+ \int_{0}^{1} \exp\left(-\frac{\alpha x}{b} + \frac{(1-\alpha)(x-1)}{b}\right) \dd x \\
&+ \int_{1}^{\infty} \exp\left(-\frac{\alpha x}{b} - \frac{(1-\alpha)(x-1)}{b}\right) \dd x \Bigg] \\
\end{aligned} \\
&=\frac{e^{1 / b}-1}{e^{1 / b}+1} \cdot \left[be^{(\alpha-1)/b} + \frac{b}{2\alpha-1}\left(e^{(\alpha-1)/b} - e^{-\alpha/b}\right) + be^{-\alpha/b}\right] \\
&=\frac{e^{1 / b}-1}{e^{1 / b}+1}\left(\frac{2\alpha b}{2\alpha-1}e^{(\alpha-1)/b} + \frac{2\alpha b - 2b}{2\alpha-1}e^{-\alpha/b}\right).
\end{align}
\end{proof}

In the corollary below, we use the notation $\tau$ to denote the privacy loss instead of the usual notation $\epsilon$ in order to differentiate it from the pure DP loss that denote by $\epsilon$.

\begin{corollary}
\label{cor:geometric-satisfies-renyi-dp}
Let $q: \mathcal{X} \rightarrow \mathbb{R}$. Then \textsc{BaseGeometric}$(q(x), \epsilon)$ satisfies $(\alpha, \tau)$-RDP privacy with respect to $x$, where
\begin{equation}
    \tau = \frac{1}{\alpha-1} \log\left[\frac{e^{\epsilon}-1}{e^{\epsilon}+1}\left(\frac{2\alpha }{\epsilon(2\alpha-1)}e^{(\alpha-1)\epsilon} + \frac{2(\alpha - 1)}{\epsilon(2\alpha-1)}e^{-\alpha\epsilon}\right)\right].
\end{equation}
\end{corollary}

\subsection{Generalized parallel composition lemmas}
\label{sec:generalized-parallel-composition}

In this section, we give generalized parallel composition lemmas for pure DP, zCDP, and RDP.
The statements we give are generalizations of the standard statements of parallel composition.

Let the \emph{maximum degree} of a set family $F = \{S_i\}$, $S_i \subseteq S$ be the maximum number of sets containing any fixed element of $S$. That is, 
\begin{equation}
    degree(F) = max_{s \in S} |\{S_i \in F | s \in S_i\}|
\end{equation}

\begin{lemma}
\label{lem:generalized-parallel-composition-dp}
Let $F = \{S_1,...,S_k\}$ be family of subsets of the input domain with maximum degree $z$. Let $M_1,\ldots,M_k$ each provide $\epsilon$-differential privacy. Then the mechanism $M(x) = (M_1(x \cap S_1), \ldots, M_k(x \cap S_k))$ provides $(z \cdot \epsilon)$-differential privacy.
\end{lemma}
\begin{proof}
Suppose $x$ and $x'$ are neighbors, and let $r$ be the (only) record in their symmetric difference.
Let $i_1, \ldots, i_j$ be the indices of the sets in $F$ containing $r$.
$j \le z$ since the maximum degree of $F$ is $z$. Then for any $y$,
\begin{align}
\Pr[M(x) = y] &= \prod_{i=1}^k \Pr[M_i(x \cap S_i) = y_i] \\
&= \prod_{i \in \{i_1,\ldots,i_j\}} \Pr[M_i(x \cap S_i) = y_i] \prod_{i\not\in \{i_1,\ldots,i_j\}} \Pr[M_i(x' \cap S_i) = y_i] \\
&\le \prod_{i \in \{i_1,\ldots,i_j\}} \left[ e^\epsilon \Pr[M_i(x' \cap S_i) = y_i]\right]  \prod_{i\not\in \{i_1,\ldots,i_j\}} \Pr[M_i(x' \cap S_i) = y_i] \\
&= e^{j\epsilon} \cdot \prod_{i=1}^k \Pr[M_i(x' \cap S_i) = y_i] \\
&= e^{j\epsilon} \cdot \Pr[M(x') = y].
\end{align}
\end{proof}

We can give an analogous lemma for zCDP. 

\begin{lemma}
\label{lem:generalized-parallel-composition-zcdp}
Let $F = \{S_1,...,S_k\}$ be family of subsets of the input domain with maximum degree $z$. Let $M_1,\ldots,M_k$ each provide $\rho$-zCDP. Then the mechanism $M(x) = (M_1(x \cap S_1), \ldots, M_k(x \cap S_k))$ provides $(z \cdot \rho)$-zCDP.
\end{lemma}

The proof of \ref{lem:generalized-parallel-composition-zcdp} requires the following property on the R\'enyi divergence, given in Lemma 2.2 of \cite{BunS16}.

\begin{lemma}(\cite{BunS16})
Let $P_{\Omega}$ and $Q_{\Omega}$ be distributions on $\Omega$, and $P_{\Theta}$ and $Q_{\Theta}$ be distributions on $\Theta$. Let $P = P_{\Omega}P_{\Theta}$ and $Q = Q_{\Omega}Q_{\Theta}$. Then
\begin{equation}
    D_{\alpha}(P \| Q) = D_{\alpha}(P_{\Omega} || Q_{\Omega}) + D_{\alpha}(P_{\Theta} || Q_{\Theta})
\end{equation}
\end{lemma}

With this, we can prove Lemma~\ref{lem:generalized-parallel-composition-zcdp}.

\begin{proof}[Proof of Lemma~\ref{lem:generalized-parallel-composition-zcdp}]
Suppose $x$ and $x'$ are neighbors, and let $r$ be the (only) record in their symmetric difference.
Let $i_1, \ldots, i_j$ be the indices of the sets in $F$ containing $r$.
$j \le z$ since the maximum degree of $F$ is $z$.

\begin{align}
D_{\alpha}(M(x) \| M(x')) &= \sum_{i=1}^k D_{\alpha}(M_i(x \cap S_i) \| M_i(x' \cap S_i)) \\
&= \sum_{i \in \{i_1,\ldots,i_j\}} D_{\alpha}(M_i(x \cap S_i) \| M_i(x' \cap S_i)) \\
&\le \sum_{i \in \{i_1,\ldots,i_j\}} \alpha \cdot \rho \\
&\le \alpha \cdot (z \cdot \rho).
\end{align}
\end{proof}
Finally, we give a generalized parallel composition lemma for RDP. The proof is nearly identical to the proof of Lemma~\ref{lem:generalized-parallel-composition-zcdp} so we omit it.

\begin{lemma}
\label{lem:generalized-parallel-composition-rdp}
Let $F = \{S_1,...,S_k\}$ be family of subsets of the input domain with maximum degree $z$. Let $M_1,\ldots,M_k$ each provide $(\alpha, \epsilon)$-RDP. Then the mechanism $M(x) = (M_1(x \cap S_1), \ldots, M_k(x \cap S_k))$ provides $(\alpha, z \cdot \epsilon)$-RDP.
\end{lemma}


\section{Notation \& Algorithms}
\begin{table}[t]
    \centering
    \begin{tabular}{c p{.8\linewidth}}
        \toprule
        \textbf{Notation} & \textbf{Description} \\
        \midrule
        $\omega$ & the number of population group levels \\
        $\mathcal{P}_i$ & population group level $i$  \\
        $\rho_i$ & the privacy loss budget allocated to population group level $i$ \\
        $g_i$ & a function mapping records to the set of population groups in $\mathcal{P}_i$ to which the record belongs \\
        $\Delta(g_i)$ & $\max_{r \in \mathcal{I}} |g_i(r)|$
    \end{tabular}
    \caption{A summary of the notation used in Section~\ref{sec:algorithm-description}}
    \label{tab:algorithm-notation}
\end{table}

\begin{algorithm}[t]
\caption{\label{alg:safetab-main-algorithm} The main \safetabp[$\Gamma$] algorithm.}
\begin{algorithmic}[1]
\Require
$df$: private dataframe with attributes [BlockID, RaceEth, Sex, Age] and one row for each person in the US
\Require $\Gamma$: Noise Mechanism that is either Geometric Mechanism or Discrete Gaussian Mechanism
\Require $\{\rho_i\}_{i \in [1,\omega]}$: Privacy parameters for each population group level $i \in [1, \omega]$.
\Require $\gamma$: The fraction of the privacy loss budget to be used in Stage 1 of the two stage tabulation algorithm.

\Procedure{SafeTab-P}{$df$, $\Gamma$, $\{\rho_i\}$, $\gamma$}

\For{$i \in [1, \omega]$} \label{line:pop-group-level-loop}
\State $df_i \leftarrow df$.flatmap($g_i$);
\Comment{$df_i$ has schema [PopGroup, Sex, Age]}
\State $s \leftarrow \Delta(g_i)$ 
\Comment{1 row in $df$ may result in $\leq s$ rows in $df_i$}
\For{$P \in \mathcal{P}_i$ \label{line:iteration-loop}}
\State $df_P$ $\leftarrow$ $df_i$.filter(PopGroup $== P$)
\State \Call{TabulatePopulationGroup}{$df_p$, $P$, $\Gamma$,  $\rho_i/s$, $\epsfrac$}
\EndFor
\EndFor
\EndProcedure
\end{algorithmic}
\end{algorithm}

\begin{algorithm}[t]
\caption{\label{alg:safetab-tabulate-pop-group} Subroutine of \safetabp to tabulate statistics for a single population group.}
\begin{algorithmic}[1]
\Require $df$: a private dataframe with attributes [BlockID, RaceEth, Sex, Age]. This dataframe should contain the records in the population group.
\Require $P$: The population group.
\Require $\Gamma$: Noise Mechanism that is either Geometric Mechanism or Discrete Gaussian Mechanism 
\Require $\rho$: Privacy loss budget for this subroutine.
\Require $\epsfrac$: Fraction of $\rho$ used in the adaptive algorithm

\Procedure{TabulatePopulationGroup}{$df, P, \Gamma, \rho, \epsfrac$}

\If{$P$ $\in$ TotalOnly}
\State \textit{// For TotalOnly population groups, only report noisy total counts}
\State \textbf{Output} \Call{NoisyCount}{$df, \Gamma, \rho$} \label{line:noisy-total-only}

\State
\Else
\State \textit{// For the rest of the population groups, adaptively choose the statistics released}
\State \textit{//  based on the noisy total count of the population group.}
\State \textit{// Step 1: Compute the noisy total count using $\rho\epsfrac$ privacy loss budget}
\State total $\leftarrow$ \Call{NoisyCount}{$df, \Gamma, \epsfrac\rho$} \label{line:stage1-count}
\Comment{Compute noisy total}
\State
\State \textit{// Step 2: Release statistics based on the noisy count with $\rho(1 - \epsfrac)$ privacy loss budget}

\If{total $< \theta_1$}
\State \textbf{Output} \Call{NoisyCount}{df, $\Gamma$, $(1-\epsfrac)\rho$}\label{line:age1} 
\Comment{Output the total}

\ElsIf{total $< \theta_2$}
\For{df\_group $\in$ df.map(Age $\rightarrow$ Age4).groupby(Sex, Age4)}
\State \textbf{Output} {\Call{NoisyCount}{df\_group, $\Gamma$, $(1-\epsfrac)\rho$}} \label{line:age4}
\Comment{Sex X Age4 marginal}
\EndFor

\ElsIf{total $< \theta_3$}
\For{df\_group $\in$ df.map(Age $\rightarrow$ Age9).groupby(Sex, Age9)}
\State \textbf{Output} {\Call{NoisyCount}{df\_group, $\Gamma$, $(1-\epsfrac)\rho$}} \label{line:age9}
\Comment{Sex X Age9 marginal}
\EndFor

\Else 
\For{df\_group $\in$ df.map(Age $\rightarrow$ Age23).groupby(Sex, Age23)}
\State \textbf{Output} {\Call{NoisyCount}{df\_group, $\Gamma$, $(1-\epsfrac)\rho$}}\label{line:age23}
\Comment{Sex X Age23 marginal}
\EndFor

\EndIf
\EndIf
\EndProcedure
\end{algorithmic}
\end{algorithm}

\begin{algorithm}[t]
\caption{\label{alg:noisy-count} Noisy Count Mechanism}
\begin{algorithmic}[1]
\Require{$df$: The dataframe.}
\Require $\Gamma$: The noise mechanism. This should be either Geometric or Discrete Gaussian.
\Require $\rho$: The privacy loss budget for this subroutine. Its interpretation depends on the chosen noise mechanism.
\Procedure{NoisyCount}{$df, \Gamma, \rho$}
\If{$\Gamma$ is Geometric}
\State \textit{// $\rho$ is the pure-DP privacy loss parameter}
\State \Return \Call{BaseGeometric}{$df$.count(), $\rho$}
\ElsIf{$\Gamma$ is Discrete Gaussian}
\State \textit{// $\rho$ is the zCDP privacy loss parameter}
\State \Return \Call{BaseDiscreteGaussian}{$df$.count(), $\rho$}
\EndIf
\EndProcedure

\end{algorithmic}
\end{algorithm}

\begin{algorithm}
\caption{\label{alg:base-geometric} The base geometric mechanism.}
\begin{algorithmic}[1]
\Require $c$: An integer.
\Require $\epsilon$: The desired privacy loss parameter.
\Procedure{BaseGeometric}{$c, \epsilon$}
\State $y \gets \mathcal{L}_{\mathbb{Z}}\left(\frac{1}{\epsilon}\right)$ 
\State \textbf{return} $c + y$
\EndProcedure
\end{algorithmic}
\end{algorithm}

\begin{algorithm}[t]
\caption{\label{alg:base-discrete-gaussian} The base discrete Gaussian mechanism.}
\begin{algorithmic}[1]
\Require $c$: An integer.
\Require $\rho$: The desired privacy loss parameter.
\Procedure{BaseDiscreteGaussian}{$c, \rho$}
\State $y \gets \mathcal{N}_{\mathbb{Z}}\left(\frac{1}{2\rho}\right)$
\State \textbf{return} $c + y$
\EndProcedure
\end{algorithmic}
\end{algorithm}

\clearpage


\section{SafeTab[Geometric] Privacy and Error Analysis} 
\label{sec:safetab-geometric-privacy}

In this section, we give two privacy analyses for the SafeTab[Geometric] algorithm, as well as an analysis of the error of the algorithm.
Both the privacy analyses in this section, as well as the privacy analysis in Section~\ref{sec:safetab-discrete-gaussian-privacy}, follow a very similar recipe.
Each result follows from the privacy properties of the base mechanism, combined with the composition rules given in sections~\ref{sec:privacy-prelim} and~\ref{sec:new-privacy}.
Because the composition results for the different privacy definitions are essentially the same, the privacy analysis proofs are all very similar.
For completeness, we give all the privacy proofs.

Note that it is also possible to give a zCDP analysis of SafeTab[Geometric]. However, this analysis is not as tight as RDP analysis so we omit it.

\subsection{Pure-DP Privacy Analysis}
\label{sec:geometric-pure-dp-analysis}
In this section, show that the SafeTab[Geometric] algorithm presented in Section~\ref{sec:algorithm-description} satisfies pure differential privacy.

Note that we chose to use $\rho$ as a parameter in the SafeTab algorithm description.
When analyzing the geometric version of the algorithm under pure DP, the parameter $\rho$ corresponds to the pure dp loss, which is generally denoted as $\epsilon$.

\begin{theorem}
\label{thm:safetab-saisfies-pure-dp}
Let $\rho_{total} = \sum_{i=1}^{\omega} \rho_i$. Algorithm~\ref{alg:safetab-main-algorithm} satisfies $\rho_{total}$-differential privacy when $\Gamma$ is the Geometric Mechanism.
\end{theorem}

\begin{proof}
The proof follows from a combination of composition rules along with the fact that the base mechanism, \textsc{BaseGeometric} satisfies pure dp.

First, we claim that the procedure \textproc{NoisyCount} where $\Gamma = $ \emph{Geometric} satisfies $\rho$-differential privacy, where $\rho$ is the privacy parameter input to \textproc{NoisyCount}.
This follows directly from Lemma~\ref{lem:geometric-satisfies-pure-dp}.

Next, we claim that the procedure \textproc{TabulatePopulationGroup} in Algorithm~\ref{alg:safetab-tabulate-pop-group} satisfies $\rho$-differential privacy with respect to the input dataframe, where $\rho$ is the privacy parameter input to the procedure.
Note that \textproc{TabulatePopulationGroup} actually uses one of two algorithms depending on whether the population group is in the set TotalOnly. We consider each of these algorithms.

\textbf{Case 1:} $P \in $ TotalOnly. In this case the procedure simply calls \textsc{NoisyCount}, which satisfies $\rho$-differential privacy.

\textbf{Case 2:} $P \not \in $ TotalOnly. In this case, the procedure can be decomposed into two parts. First, we call \textsc{NoisyCount} with a budget of $\epsfrac\rho$. Then, we use the result to group the data by sex and age, and for each group we make a call to \textsc{NoisyCount} with a budget of $(1-\epsfrac)\rho$.
The composition of the calls on all the groups satisfies $(1-\epsfrac)\rho$ by Lemma~\ref{lem:generalized-parallel-composition-dp}.
The (adaptive) composition of the two parts has total privacy loss $\rho$ by Lemma~\ref{lem:sequential-composition-dp}.

Next, we claim that the $i$th loop of the \textbf{for} loop on line~\ref{line:pop-group-level-loop} of Algorithm~\ref{alg:safetab-main-algorithm} satisfies $\rho_i$-differential privacy.
By the definition of $s$, any particular record can appear in the input ($df_P$) of at most $s$ calls to \textsc{TabulatePopulationGroup}.
Therefore, by Lemma~\ref{lem:generalized-parallel-composition-dp}, the total privacy loss of the loop is $s$ times the privacy loss of \textsc{TabulatePopulationGroup}, i.e. $s \cdot \frac{\rho_i}{s} = \rho_i$.

Finally, the overall algorithm satisfies $(\sum_{i=1}^{\omega} \rho_i)$-differential privacy by Lemma~\ref{lem:sequential-composition-dp}.
\end{proof}

\subsubsection{A note on the generalized parallel composition lemma}

Rather than using the generalized parallel composition lemma (Lemma~\ref{lem:generalized-parallel-composition-dp}) to analyze the algorithm, we could have used the popular \emph{stability}-based accounting method \cite{McSherry09}.
This is an approach for calculating the sensitivity of a query made up of multiple database transformations followed by a measurement.
This method gives a tight analysis under differential privacy using the geometric mechanism because combining a sensitivity $x$ query and a sensitivity $y$ query into a single query (outputting a vector) has sensitivity $x+y$.
Therefore, analyzing the algorithm as a few high sensitivity queries (stability analysis) is equivalent to analyzing it as the the composition of many low sensitivity queries.

On the other hand, adding noise from the discrete Gaussian distribution to satisfy zCDP requires the scale of the noise to be proportional to the square of the sensitivity, rather than the sensitivity.
This means that a single query with sensitivity $x+y$ will require significantly more noise than the composition of mechanisms that answer sensitivity $x$ and sensitivity $y$ queries respectively.
Therefore, we can get a tighter analysis by analyzing the algorithm as the composition of many low sensitivity queries (sensitivity 1 in the case of SafeTab) rather than fewer high sensitivity queries. This requires using the generalized parallel composition instead of stability accounting.

\subsection{RDP Privacy Analysis}
\label{sec:geometric-rdp-analysis}

In this section, we show that the SafeTab[Geometric] algorithm presented in Section~\ref{sec:algorithm-description} satisfies RDP.

\begin{theorem}
\label{thm:safetab-saisfies-rdp}
Let $\tau(\alpha, \epsilon)$ denote the RDP privacy bound for \textsc{BaseGeometric} given in Corollary~\ref{cor:geometric-satisfies-renyi-dp}. That is,
\begin{equation}
  \label{eq:1}
    \tau(\alpha, \epsilon) = \frac{1}{\alpha-1} \log\left[\frac{e^{\epsilon}-1}{e^{\epsilon}+1}\left(\frac{2\alpha }{\epsilon(2\alpha-1)}e^{(\alpha-1)\epsilon} + \frac{2(\alpha - 1)}{\epsilon(2\alpha-1)}e^{-\alpha\epsilon}\right)\right].
\end{equation}
Algorithm~\ref{alg:safetab-main-algorithm} satisfies
\begin{equation}
\left(\alpha,\sum_{i=1}^{\omega}\left[ s \cdot \max\left[\tau\left(\alpha, \frac{\gamma\rho_{i}}{s}\right) + \tau\left(\alpha, \frac{(1 - \gamma)\rho_{i}}{s}\right), \tau\left(\alpha, \frac{\rho_{i}}{s}\right)\right]\right]\right) \text{-RDP}
\end{equation}
when $\Gamma$ is the Geometric Mechanism.
\end{theorem}

\begin{proof}
The proof follows from a combination of composition rules along with the fact that the base mechanism, \textsc{BaseGeometric} satisfies $(\alpha, \tau(\alpha, \epsilon))$-RDP.

First, we claim that the procedure \textproc{NoisyCount} where $\Gamma = $ \emph{Geometric} satisfies $(\alpha, \tau(\alpha, \rho))$-RDP, where $\rho$ is the privacy parameter input to \textproc{NoisyCount}.
This follows directly from Corollary~\ref{cor:geometric-satisfies-renyi-dp}.

Next, we claim that the procedure \textproc{TabulatePopulationGroup} in Algorithm~\ref{alg:safetab-tabulate-pop-group} satisfies $(\alpha, \max(\tau(\alpha, \gamma\rho) + \tau(\alpha, (1 - \gamma)\rho), \tau(\alpha, \rho)))$-RDP with respect to the input dataframe, where $\rho$ is the privacy parameter input to the procedure.
Note that \textproc{TabulatePopulationGroup} actually uses one of two algorithms depending on whether the population group is in the set TotalOnly. We consider each of these algorithms.

\textbf{Case 1:} $P \in $ TotalOnly. In this case the procedure simply calls \textsc{NoisyCount}, which satisfies $(\alpha, \tau(\alpha, \rho))$-RDP.

\textbf{Case 2:} $P \not \in $ TotalOnly. In this case, the procedure can be decomposed into two parts. First, we call \textsc{NoisyCount} with a budget of $\epsfrac\rho$. Then, we use the result to group the data by sex and age, and for each group we make a call to \textsc{NoisyCount} with a budget of $(1-\epsfrac)\rho$.
The composition of the calls on all the groups satisfies $(\alpha, \tau(\alpha, (1-\epsfrac)\rho))$-RDP by Lemma~\ref{lem:generalized-parallel-composition-rdp}.
The (adaptive) composition of the two parts satisfies $(\alpha, \tau(\alpha, \gamma\rho) + \tau(\alpha, (1 - \gamma)\rho))$-RDP  by Lemma~\ref{lem:sequential-composition-rdp}.

Next, we claim that the $i$th loop of the \textbf{for} loop on line~\ref{line:pop-group-level-loop} of Algorithm~\ref{alg:safetab-main-algorithm} satisfies
\begin{equation}
\left(\alpha, s \cdot \max\left[\tau\left(\alpha, \frac{\gamma\rho_{i}}{s}\right) + \tau\left(\alpha, \frac{(1 - \gamma)\rho_{i}}{s}\right), \tau\left(\alpha, \frac{\rho_{i}}{s}\right)\right]\right) \text{-RDP}.
\end{equation}
By the definition of $s$, any particular record can appear in the input ($df_P$) of at most $s$ calls to \textsc{TabulatePopulationGroup}.
Therefore, by Lemma~\ref{lem:generalized-parallel-composition-rdp}, the total privacy loss of the loop is $s$ times the privacy loss of \textsc{TabulatePopulationGroup}.

Finally, the overall algorithm satisfies

\begin{equation}
\left(\alpha,\sum_{i=1}^{\omega}\left[ s \cdot \max\left[\tau\left(\alpha, \frac{\gamma\rho_{i}}{s}\right) + \tau\left(\alpha, \frac{(1 - \gamma)\rho_{i}}{s}\right), \tau\left(\alpha, \frac{\rho_{i}}{s}\right)\right]\right]\right) \text{-RDP}
\end{equation}
by Lemma~\ref{lem:sequential-composition-rdp}.

\end{proof}

\subsection{Error Bounds}\label{sec:geometric-error-bound}
We next examine the utility of Algorithm \ref{alg:safetab-main-algorithm} with base geometric mechanism. We first restate a portion of Lemma 30 from \cite{CanonneK2020}:

\begin{lemma}\label{lem:geometric-tail-bounds}
Let $b > 0$ and let $Y \leftarrow \mathcal{L}_{\mathbb{Z}}(b)$. For all $y \in \mathbb{R}$,
\begin{equation}
    \Pr[Y \geq y] = \Pr[Y \leq -y] \leq \frac{e^{-\frac{\lceil y \rceil}{b}}}{1 + e^{-\frac{1}{b}}}.
\end{equation}


\end{lemma}


Hence for all $y \in \mathbb{R}$, 
\begin{equation}
P[Y > y] = P[Y < -y] \leq \frac{e^{-\frac{\lfloor y \rfloor}{b}}}{1 + e^{-\frac{1}{b}}} - \frac{e^{1/b}-1}{e^{1/b}+1} \cdot e^{-\lfloor y \rfloor /b} = 
\frac{e^{-\lfloor y \rfloor /b}}{1 + e^{1/b}}. 
\end{equation}
It follows that for any $y \in \mathbb{R}$, 
\begin{equation}
     Y \in \left[y - \left\lfloor b\ln{\left(\frac{2}{p\left(1 + e^{\frac{1}{b}}\right)}\right)}\right\rfloor,y + \left\lfloor b\ln{\left(\frac{2}{p\left(1 + e^{\frac{1}{b}}\right)}\right)}\right\rfloor\right].
\end{equation}
with probability $1 - p$. Hence, the margin of error of a 95\% confidence interval is
\begin{equation}
    \left\lfloor b\ln{\left(\frac{40}{1 + e^{\frac{1}{b}}}\right)} \right\rfloor.
\end{equation}
Note that for a fixed integral 95\% MOE, we have $b \in \left(\frac{MOE}{\ln 20}, \frac{MOE + 1}{\ln 20}\right)$.

\begin{corollary}\label{cor:geometric-scale-from-moe}
The base geometric mechanism run with parameter $\epsilon = \frac{\ln(20)}{\lfloor MOE \rfloor + 1}$ has a 95\% margin on error of at most $MOE$.
\end{corollary}

Applying the error bounds, the total count estimate of a TotalOnly population group in population group level $i$ would have a margin of error of $\left\lfloor \frac{s}{\rho_i}\ln{\left(\frac{40}{1+e^{\rho_i/s}}\right)}\right\rfloor$. For a non-TotalOnly population group in level $i$, the margin of error in a single sex by age group is  $\left\lfloor\frac{s}{(1-\gamma)\rho_i}\ln{\left(\frac{40}{1+e^{(1-\gamma)\rho_i/s}}\right)}\right\rfloor$.

\section{SafeTab[Discrete Gaussian] Privacy Analysis}
\label{sec:safetab-discrete-gaussian-privacy}

In this section, we give a zCDP analysis of the SafeTab[Discrete Gaussian] algorithm, as well as an analysis of the error of the algorithm.
This analysis follows the same formula as the analyses in Section~\ref{sec:safetab-geometric-privacy}.

\subsection{zCDP privacy analysis}
\label{sec:gaussian-zcdp-analysis}

In this section, show that the SafeTab[Discrete Gaussian] algorithm presented in Section~\ref{sec:algorithm-description} satisfies zero-concentrated differential privacy (zCDP).

\begin{theorem}
\label{thm:safetab-satisfies-zcdp}
Let $\rho_{total} = \sum_{i=1}^{\omega} \rho_i$. Algorithm~\ref{alg:safetab-main-algorithm} satisfies $\rho_{total}$-zCDP when $\Gamma$ is the Discrete Gaussian Mechanism.
\end{theorem}
\begin{proof}
The proof of Theorem~\ref{thm:safetab-satisfies-zcdp} via the combination of composition rules along with the fact that the base mechanism, \textsc{BaseDiscreteGaussian}.

First, we claim that the procedure \textproc{NoisyCount} where $\Gamma = $ \emph{Discrete Gaussian} satisfies $\rho$-zCDP, where $\rho$ is the privacy parameter input to \textproc{NoisyCount}.
This follows directly from Lemma~\ref{lem:discrete-gaussian-satisfies-zcdp}.

Next, we claim that the procedure \textproc{TabulatePopulationGroup} in Algorithm~\ref{alg:safetab-tabulate-pop-group} satisfies $\rho$-zCDP with respect to the input dataframe, where $\rho$ is the privacy parameter input to the procedure.
Note that \textproc{TabulatePopulationGroup} actually uses one of two algorithms depending on whether the population group is in the set TotalOnly. We consider each of these algorithms.

\textbf{Case 1:} $P \in $ TotalOnly. In this case the procedure simply calls \textsc{NoisyCount}, which satisfies $\rho$-zCDP.

\textbf{Case 2:} $P \not \in $ TotalOnly. In this case, the procedure can be decomposed into two parts. First, we call \textsc{NoisyCount} with a budget of $\epsfrac\rho$. Then, we use the result to group the data by sex and age, and for each group we make a call to \textsc{NoisyCount} with a budget of $(1-\epsfrac)\rho$.
The composition of the calls on all the groups satisfies $(1-\epsfrac)\rho$ by Lemma~\ref{lem:generalized-parallel-composition-zcdp}.
The (adaptive) composition of the two parts has total privacy loss $\rho$ by Lemma~\ref{lem:sequential-composition-zcdp}.

Next, we claim that the $i$th loop of the \textbf{for} loop on line~\ref{line:pop-group-level-loop} of Algorithm~\ref{alg:safetab-main-algorithm} satisfies $\rho_i$-zCDP.
By the definition of $s$, any particular record can appear in the input ($df_P$) of at most $s$ calls to \textsc{TabulatePopulationGroup}.
Therefore, by Lemma~\ref{lem:generalized-parallel-composition-zcdp}, the total privacy loss of the loop is $s$ times the privacy loss of \textsc{TabulatePopulationGroup}, i.e. $s \cdot \frac{\rho_i}{s} = \rho_i$.

Finally, the overall algorthm satisfies $(\sum_{i=1}^{\omega} \rho_i)$-zCDP by Lemma~\ref{lem:sequential-composition-zcdp}.
\end{proof}

\subsection{Error bounds}\label{sec:discrete-gauss-error-bound}
We next examine the utility of Algorithm \ref{alg:safetab-main-algorithm} with discrete Gaussian noise. We begin by stating a portion of Proposition 25 from \cite{CanonneK2020}.

\begin{proposition}[Proposition 25 in \cite{CanonneK2020}]\label{lem:discrete-gaussian-bound}

For all $m \in \mathbb{Z}$ with $m \geq 1$, and for all $\sigma \in \mathbb{R}$ with $\sigma > 0$, $\Pr[X \geq m]_{X \leftarrow \mathcal{N}_{\mathbb{Z}}(\sigma^2)} \leq \Pr[X \geq m-1]_{X \leftarrow \mathcal{N}(\sigma^2)}$.

\end{proposition}

The following corollary is immediate. 
\begin{corollary}
For all $m, \sigma \in \mathbb{R}$ with $x \geq 1$ and $\sigma > 0$,  $\Pr[X > x]_{X \leftarrow \mathcal{N}_{\mathbb{Z}}(\sigma^2)} \leq \Pr[X > \lfloor x \rfloor]_{X \leftarrow \mathcal{N}(\sigma^2)}$.
\end{corollary}

Figure 2 of \cite{CanonneK2020} provides an intuitive visualization of these tail bounds.
It follows that for all $x \in \mathbb{R}$ with $x \geq 1$, $X \in [ x  - \lfloor 1.96\sigma \rfloor,  x + \lfloor 1.96\sigma \rfloor]$ with probability at least 95\%.
That is, the 95\% margin of error is given by $\lfloor 1.96\sigma \rfloor$.

\eat{
\begin{proof}
\begin{equation}
    \Pr[X > x]_{X \leftarrow \mathcal{N}_{\mathbb{Z}}(\sigma^2)} = \frac{\sum_{k=\lfloor x + 1 \rfloor}^\infty  e^{-x^2/2\sigma^2}}{\sum_{z \in \mathbb{Z}} e^{-z^2/2\sigma^2}}
\end{equation}

The denominator is at least $\sqrt{2\pi\sigma^2}$ by Fact 19 of the paper. For the numerator,

\begin{equation}
    \sum_{k=\lfloor x + 1 \rfloor}^\infty  e^{-x^2/2\sigma^2} = \cdots = \sqrt{2\pi\sigma^2} \Pr[X > \lfloor x + 1 \rfloor -1] = \sqrt{2\pi\sigma^2}\Pr[X > \lfloor x \rfloor ] 
\end{equation}

\end{proof}
}


Hence for a population group in level $i$ in the TotalOnly set, the margin of error in the directly computed total estimate from line \ref{line:noisy-total-only} in Algorithm \ref{alg:safetab-tabulate-pop-group}  is $\left \lfloor1.96\sqrt{\frac{s}{2\rho_i}} \right \rfloor$. For the population groups in level $i$ not in the TotalOnly set, the margin of error in a single sex by age group in Algorithm \ref{alg:safetab-tabulate-pop-group} is 
$\left \lfloor1.96\sqrt{\frac{s}{2(1-\gamma)\rho_i}} \right \rfloor$.

\begin{corollary}\label{cor:dgauss-rho-from-moe}
The base discrete Gaussian mechanism run with $\rho = \frac{1.92}{\lfloor MOE \rfloor^2}$ has a 95\% margin on error of at most $MOE$.
\end{corollary}

\section{Comparing SafeTab[Geomteric] vs SafeTab[Discrete Gaussian]}
\label{sec:compare}

In this section we compare SafeTab[Discrete Gaussian] and SafeTab[Geometric]. We set parameters to recommended settings, and compare the approximate dp privacy loss of the algorithms using various analyses. In particular, we fix a target margin of error for both versions of the algorithm and compute the approximate differential privacy loss for a fixed $\delta$.

\subsection{Fixing algorithm parameters}

To evaluate the privacy losses of the algorithm under approximate differential privacy, we set parameter values to pre-decisional settings suggested by subject matter experts at the Census. Parameters are set as follows.
\begin{itemize}
    \item The approximate differential privacy parameter $\delta$ is $10^{-10}$.
    \item The number of population group levels is 7. These population groups are enumerated in the \emph{Population Group Level} column of Table~\ref{tab:moe-targets}.
    \item The population group mapping function $g_i$ has stability $\Delta(g_i) = 9$ for all $i$.
    \item The parameter $\gamma$ is $0.1$.
    \item The MOE targets are given in Table~\ref{tab:moe-targets}. Note that MOE targets are set for Step 2 of the 2-step algorithm (i.e. lines~\ref{line:age1} to~\ref{line:age23} of Algorithm~\ref{alg:safetab-tabulate-pop-group}).
\end{itemize}

\begin{table}[t]
    \centering
    \begin{tabular}{c c c c c c}
    \toprule
    Population Group Level & MOE Target &  
    \multicolumn{2}{c}{Geometric ($\epsilon$)} & \multicolumn{2}{c}{Discrete Gaussian ($\rho$)}
    \\
    \cmidrule(lr{.75em}){3-4}
    \cmidrule(lr{.75em}){5-6}
    & & Step 2 & Total & Step 2 & Total \\
    \midrule
    (Nation, Detailed) & 6 & 3.84 & 4.27 & 0.481 & 0.534\\
    (State, Detailed) & 6 & 3.84 & 4.27 & 0.481 & 0.534\\
    (County, Detailed) & 11 & 2.24 & 2.49 & 0.143 & 0.159 \\
    (AIANNH, Detailed) & 11 & 2.24 & 2.49 & 0.143 &0.159\\
    (Nation, Regional) & 50 & 0.531 & 0.59 & 0.007 & 0.008\\
    (State, Regional) & 50 & 0.531 & 0.59 & 0.007 &0.008\\
    (County, Regional) & 50 & 0.531 & 0.59 & 0.007 &0.008\\
    \bottomrule
    \end{tabular}
   \caption{MOE targets for the statistics released (in Step 2 of the adaptive algorithm) at different population group levels along with the corresponding privacy loss ($\epsilon$-DP for Geometric and $\rho$-zCDP for Discrete Gaussian). The privacy loss is reported for the Step 2 (to match the MOE) as well as the total loss for that level. Step 2 loss is 90\% of Total loss at each population group level. Note that the privacy losses reported here have already been aggregated over all the population groups at the given level, so the \emph{Total} column represents the privacy loss input parameters of the SafeTab algorithm.}
   \label{tab:moe-targets}
\end{table}
\eat{\begin{table}
    \centering
    \begin{tabular}{l l l}
    \toprule
         Population Group Level & Algorithm Stage & MOE Target \\
         \midrule
         1, 2 & Total Only & 5 \\
         & Stage 1 & 50 \\
         & Stage 2 & 6 \\
         3, 4 & Stage 1 & 100 \\
         & Stage 2 & 11 \\
         5, 6, 7 & Stage 1 & 500 \\
         & Stage 2 & 50 \\
         \bottomrule
    \end{tabular}
    \caption{MOE targets for the different population group levels and algorithm stages. The algorithm stages are the counts taken at the following lines in Algorithm~\ref{alg:safetab-tabulate-pop-group}: Total Only: the noisy count on line~\ref{line:noisy-total-only}, Stage 1: the noisy count on line~\ref{line:stage1-count}, Stage 2: the noisy counts on one of lines~\ref{line:age1},~\ref{line:age4},~\ref{line:age9}, or~\ref{line:age23}. There is no \emph{Total Only} target for population group levels 3 to 7 because these levels have no population groups in the \emph{Total Only} set.}\label{tab:moe-targets}
\end{table}
}

We next convert the MOE targets into privacy parameter algorithm inputs as follows.
First, we convert the MOE target for each base mechanism into the corresponding privacy parameter for the mechanism using the tail bounds presented in the previous section.
That is, the parameter $\rho$ for the base discrete Gaussian mechanism is given by
\begin{equation}
  \label{equ:discrete-guassian-parameter-from-moe}
  \rho = \frac{1.92}{\lfloor MOE \rfloor^2},
\end{equation}
where $MOE$ is the target MOE for the base mechanism.
See Section~\ref{sec:discrete-gauss-error-bound} for the details of the MOE derivation.

The parameter $\epsilon$ for the base geometric mechanism is
\begin{equation}
  \label{equ:geometric-parameter-from-moe}
  \epsilon = \frac{\ln(20)}{\lfloor MOE \rfloor + 1}.
\end{equation}
See Section~\ref{sec:geometric-error-bound} for the details of the MOE derivation.

We next compute the parameter $\rho$ that should be given as input to \textsc{TabulatePopulationGroup}.
Since our MOE target is for step 2 of the two step algorithm, this parameter $\rho$ is  given by $\rho = (1/\gamma) \rho_{base}$ where $\rho_{base}$ is the privacy parameter we computed for the base mechanism using the MOE target.

Finally, we compute each of $\rho_{1}, \ldots, \rho_{7}$ by multiplying the $\rho$ parameter from each call to \textsc{TabulatePopulationGroup} by $s=9$.

We summarize the parameters for the two versions of the SafeTab alongside the MOE targets in in Table~\ref{tab:moe-targets}. In this table, we also calculate the total privacy loss of step 2 tabulation over all population groups at a population group level.
This loss is 90\% of the total loss.

Note that while the inputs to the SafeTab[Geometric] and SafeTab[Discrete Gaussian] are the pure dp and zCDP privacy losses respectively, it is also possible to analyze the algorithms under different privacy guarantees (e.g. in the next section we analyze the geometric version of the algorithm using RDP).

\subsection{Privacy loss comparison approach}

In this section we describe multiple analyses of the privacy loss of SafeTab[Discrete Gaussian] and SafeTab[Geometric].
Results appear Table~\ref{tab:tpdp-privacy-loss-results}.
For each analysis except the pure differential privacy analysis (where $\delta=0$), we use the approximate differential privacy loss with $\delta = 10^{-10}$.

\paragraph{Pure DP loss of SafeTab[Geometric]}

To compute the pure differential privacy loss for SafeTab[Geometric], we can simply take the sum of $\rho_{1}, \ldots \rho_{7}$ by the analysis in Section~\ref{sec:geometric-pure-dp-analysis}.

\paragraph{Approximate DP loss of SafeTab[Geometric] using RDP analysis}

From the analysis in Section~\ref{sec:geometric-rdp-analysis}, we know that the SafeTab[Geometric] algorithm satisfies $(\alpha, f(\alpha))$-RDP, where
\begin{equation}
  f(\alpha) = \sum_{i=1}^{\omega}\left[ s \cdot \max\left[\tau\left(\alpha, \frac{\gamma\rho_{i}}{s}\right) + \tau\left(\alpha, \frac{(1 - \gamma)\rho_{i}}{s}\right), \tau\left(\alpha, \frac{\rho_{i}}{s}\right)\right]\right]
\end{equation}

We next convert this guarantee to an approximate DP guarantee.
Using Lemma~\ref{lem:renyi-divergence-to-approx-dp-tight}, we have that
\begin{equation}
    \epsilon = f(\alpha) + \frac{\log(1/\delta) + (\alpha-1)\log(1 - 1/\alpha) - \log(\alpha)}{\alpha-1}.
\end{equation}

Setting $\delta = 10^{-10}$, we can find the optimal value of $\epsilon$ by minimizing the right hand expression over $\alpha \in (0, \infty)$.
This minimum is challenging to find analytically we compute the expression for $\alpha \in \{1.01, 1.02, \ldots, 9.99, 10.0\}$ and take the minimum.

\paragraph{Approximate DP loss of SafeTab[Discrete Gaussian] using zCDP experimental analysis}

From the analysis in Section~\ref{sec:gaussian-zcdp-analysis}, SafeTab[Discrete Gaussian] satisfies $\rho$-zCDP where $\rho = \sum_{i=1}^{7} \rho_{i}$. We can convert this into an approximate DP guarantee using equation~\ref{equ:zcdp-to-approx-dp-tight}. Rather than computing the infimum analytically, we compute the expression for $\alpha \in \{1.01, 1.02, \ldots, 9.99, 10.0\}$ and take the minimum.

\paragraph{Approximate DP loss of SafeTab[Discrete Gaussian] using zCDP analytic analysis}

In addition to the experimental analysis, we also convert the $\rho$-zCDP to approximate DP analytically using Lemma~\ref{lem:zcdp-to-approx-dp-loose}.

\eat{
\subsection{Results \& Discussion}
\begin{table}
  \centering
  \begin{tabular}{r l l l l}
    \toprule
    \textbf{Base Mechanism} & \multicolumn{2}{c}{Geometric} & \multicolumn{2}{c}{Discrete Gaussian} \\
    \cmidrule(lr){2-3} \cmidrule(lr){4-5}
    \textbf{Analysis} & Pure DP & RDP & zCDP (analytical) & zCDP (experimental) \\
    \midrule
    \textbf{$(\epsilon, 10^{-10})$ privacy loss} & 15.3 ($\delta = 0$) & 13.2 & 12.8 & 12.2 \\
    \bottomrule
  \end{tabular}
  \caption{The approximate differential privacy loss of the SafeTab[Geometric] and SafeTab[Discrete Gaussian] algorithm for various analyses. For all but the Pure DP analysis, $\delta=10^{-10}$. For the Pure DP analysis, $\delta=0$.}
  \label{tab:privacy-loss-results}
\end{table}
\paragraph{Privacy losses of Geometric vs Discrete Gaussian:}
Results appear in Table~\ref{tab:privacy-loss-results}. We observe the following key findings: 
\begin{itemize}
    \item The pure DP privacy loss of SafeTab[Geometric] is bounded by $\epsilon=15.3$. This is smaller than the $\epsilon=18$ that was analyzed by the Census POP team using the SafeTEx analysis tool. The lower privacy loss is due to two reasons: 
    \begin{enumerate}
        \item  The original SafeTEx analysis was done with a target MOE of 5.6 for Nation and State detailed population groups. We also updated the MOE to 6 since noise is integral and therefore it makes sense to use an integral MOE. 
        \item The original SafeTEx analysis was done assuming SafeTab used the base Laplace mechanism rather than the base geometric mechanism.
        The tail probability of the Geometric distribution are tighter than those of the Laplace distribution at integer points. This results in a smaller privacy loss.
    \end{enumerate}
    \item As expected SafeTab[Geometric] permits a smaller privacy loss ($\epsilon$) under approximate DP with ($\delta = 10^{-10}$). We are able to achieve this by analyzing SafeTab[Geometric] under R\'enyi DP. 
    \item We observe that the privacy loss of SafeTab[Discrete Gaussian] is smaller than that of SafeTab[Geometric] (when $\delta=10^{-10}$). The improvement in privacy loss is small (3\% for the analytical bound and 7\% for the experimental bound). 
\end{itemize}

\begin{table}[t]
  \centering
  \begin{tabular}{ c l l l l}
    \toprule  
    MOE for (Nation, Detailed) & \multicolumn{2}{c}{Geometric} & \multicolumn{2}{c}{Discrete Gaussian} \\
    \cmidrule(lr){2-3} \cmidrule(lr){4-5}
    and (State, Detailed) & Pure DP & RDP & zCDP (analytical) & zCDP (experimental) \\
    \midrule
     5 & 16.7 ($\delta = 0$) & 14.6 & 15.0 & 14.3 \\
     6 & 15.3 ($\delta = 0$) & 13.2 & 12.8 & 12.2 \\
     7 & 14.2 ($\delta = 0$) & 12.1 & 11.3 & 10.7 \\
     8 & 13.4 ($\delta = 0$) & 11.3 & 10.2 & 9.7 \\
     9 & 12.7 ($\delta = 0$) & 10.7 & 9.5 & 9.0 \\
     10 & 12.2 ($\delta = 0$) & 10.2 & 8.9 & 8.4  \\
     11 & 11.7 ($\delta = 0$) & 9.7 & 8.4 & 8.0 \\
    \bottomrule
  \end{tabular}
  \caption{The approximate differential privacy loss of the SafeTab[Geometric] and SafeTab[Discrete Gaussian] algorithm when $\delta=10^{-10}$ for alternate MOE values for the (Nation, Detailed) and (State, Detailed) population group levels. MOEs for all other population group levels are fixed as in Table~\ref{tab:moe-targets}.}
  \label{tab:alternate-moes}
\end{table}

\paragraph{Granularity of Statistics under Geometric vs Discrete Gaussian:}
The granularity of statistics released by SafeTab depends on the thresholds used as well as the noise scale used in Step 1 of the adaptive part of the algorithm. Under the MOE settings in Table~\ref{tab:moe-targets} the noise scales for Step 1 should be roughly the same under both SafeTab[Geometric] and SafeTab[Discrete Gaussian]. So we should expect statistics to be released at roughly the same granularity. A quantitative analysis requires running these algorithms on either the simulated or 2010 data, which we defer to future work. 

\paragraph{Alternate MOEs for Nation and State Detailed counts:}
It is evident from Table~\ref{tab:moe-targets} that most of the privacy loss results from releasing detailed counts for Nation and State population groups. We present in Table~\ref{tab:alternate-moes} the overall privacy loss that might result from changing the MOE slightly for (Nation, detailed) and (State, detailed) population group levels from 6 to 5, 7, 8, 9, 10 and 11. We did not increase beyond 11, as that is the MOE for County and AIANNH detailed population groups. We note that  increasing the MOE has a significant impact on the privacy loss. The zCDP (experimental) privacy loss bounds drops from around 12 to around 8, about a third reduction in the privacy loss. 
}


\end{document}